\documentclass[aps,prx,twocolumn,superscriptaddress]{revtex4-2}
\usepackage{amssymb}
\usepackage{amsmath,bm}
\usepackage{graphicx,color}
\usepackage{bbm}
\usepackage{newtxmath}
\usepackage{multirow}
\usepackage{xcolor}
\usepackage{hyperref}

\newcommand{\A}{\mathcal{A}}
\newcommand{\rout}{r_\mathrm{out}}
\newcommand{\rin}{r_\mathrm{in}}

\newcommand{\uel}{u^\mathrm{el}}
\newcommand{\sigmael}{\sigma_\mathrm{el}}
\newcommand{\dif}{\mathrm{d}}

\newcommand{\MM}[1]{\textcolor{black}{#1}}

\begin{document}

\title{Odd dipole screening in disordered matter}

\author{Yael Cohen}
\affiliation{Racah Institute of Physics The Hebrew University of Jerusalem, Jerusalem, 9190401, Israel}
\author{Amit Schiller}
\affiliation{Racah Institute of Physics The Hebrew University of Jerusalem, Jerusalem, 9190401, Israel}
\author{Dong Wang}
\affiliation{Department of Mechanical Engineering and Materials Science, Yale University, New Haven, 06520, Connecticut, USA}
\author{Joshua A. Dijksman}
\affiliation{Van der Waals-Zeeman Institute, Institute of Physics, University of Amsterdam, Science Park 904, Amsterdam, 1098XH, The Netherlands}
\author{Michael Moshe}
\email{michael.moshe@mail.huji.ac.il}
\affiliation{Racah Institute of Physics The Hebrew University of Jerusalem, Jerusalem, 9190401, Israel}

\begin{abstract}
Disordered solids, straddling the solid-fluid boundary, lack a comprehensive continuum mechanical description. They exhibit a complex microstructure wherein multiple meta-stable states exist. Deforming disordered solids induces particles rearrangements enabling the system to transition between meta-stable states. A dramatic consequence of these transitions is that quasistatic deformation cycles 
modify the reference state, facilitating the storage and release of mechanical energy.
Here we develop a  continuum mechanical theory of disordered solids, which accounts for the absence of a reference state and the lack of conserved potential energy.
Our theory, which introduces a new modulus 
describing non-conservative mechanical screening, reduces to classical elasticity in the absence of screening.    
We analytically derive predictions for the deformation field for various perturbations and geometries. While our theory applies to general disordered solids, we focus on a two-dimensional disordered granular system and predict accurately the non-affine displacement fields observed in experiments for both small and large deformations, along with the observable vanishing shear modulus.
The new proposed moduli satisfy universal relations that are independent of the specific experimental realization.
Our work thus forms the basis of an entirely new family of continuum descriptions of the mechanics of disordered solids.
\end{abstract}
	

\maketitle

\section{Introduction}
Solid materials are capable of supporting external stresses, while liquids flow indefinitely in response to similar external perturbations. This clear distinction between the mechanical behaviors of different phases of matter is disrupted by disordered solids. Examples such as granular and glassy materials exhibit a combination of fluid-like and solid-like properties simultaneously, \MM{with the onset of mechanical rigidity being protocol dependent}   \cite{90Camp, 92JN, 96JNB, 13AFP}.
Shear perturbation uncovers the complex and distinctive mechanics of such disordered materials. This includes the phenomena of shear-jamming \cite{11BZCB,16PMJ,18DJDZB,18BC}, material failure via shear-banding \cite{74Chen, 79Argon,81Spaepen, 01BLSLG, 10Schall, 13Greer}, and non-affine anomalous responses \cite{07Tordesillas,08BB,zaccone2011approximate,14ZDB}.  
It is generally accepted that these complex phenomena are governed by the interactions between localized plastic events, realized as local material rearrangement \cite{04ML,06ML,98Falk}. \MM{In stress-supporting states,} their induced elastic-like fields are of quadrupolar nature and are modeled as Eshelby inclusions \cite{13DHP}, force dipoles \cite{20DeGiuli} and more \cite{FalkLanger98}.

\MM{Material rearrangements} have two primary consequences. First, they modify the reference state to relieve strains \cite{SollichLequeux97, livne2023geometric}. Second, they transition the system from one stable state to another,
suggesting the presence of multiple meta-stable states  \cite{14CKPUZ,sun2015granular}. Because different meta-stable states may possess distinct energies, the inclusion of such effect in a continuum theory would lead to a violation of energy conservation.
More specifically, in addition to dissipating mechanical energy as heat, \MM{material rearrangements} introduce another mechanism for releasing or storing mechanical energy, by plastically modifying the reference state, 
a mechanism that is at the heart of our work. 
While there has been a substantial body of work focused on plasticity in disordered matter \cite{ argon1979plastic, taub1980kinetics, FalkLanger98}, and despite an ongoing effort to develop a continuum theory of disordered materials \cite{nampoothiri2022tensor}, 
current continuum theories of matter fall short in accurately predicting the intricate anomalous deformation fields observed in disordered solids.

Here we present a novel continuum theory that describes disordered matter via the introduction of a new \MM{mechanical} constant that captures the amount mechanical work involved in a distribution of \MM{particles rearrangements}. The theory generalizes classical elasticity by incorporating the primary implications of \MM{particles rearrangements} by capturing both the alteration of the rest state, and the violation of energy conservation due to multi-stability. We test \MM{the predictions of our theory against experiments and find excellent quantitative agreement, both for the observed elastic moduli and the detailed measure deformation fields}.
	
We draw inspiration from \MM{non-hermitian physics, where the introduction of anti-symmetric coupling constants may violate the conservation of energy via breakdown of time-reversal symmetry. Examples for such theories are non-hermitian dielectrics \cite{friedland1980geometric}, odd-viscosity \cite{avron1998odd}, and recently also odd-elasticity \cite{scheibner2020odd}. 
Contrary to these works, here we study passive disordered systems by introducing a new odd term in our proposed  mechanical screening tensor. We show that odd-dipole screening accurately predicts the intricate mechanics of disordered solids. } 
\begin{figure*}
    \centering
    \includegraphics[width=\linewidth]{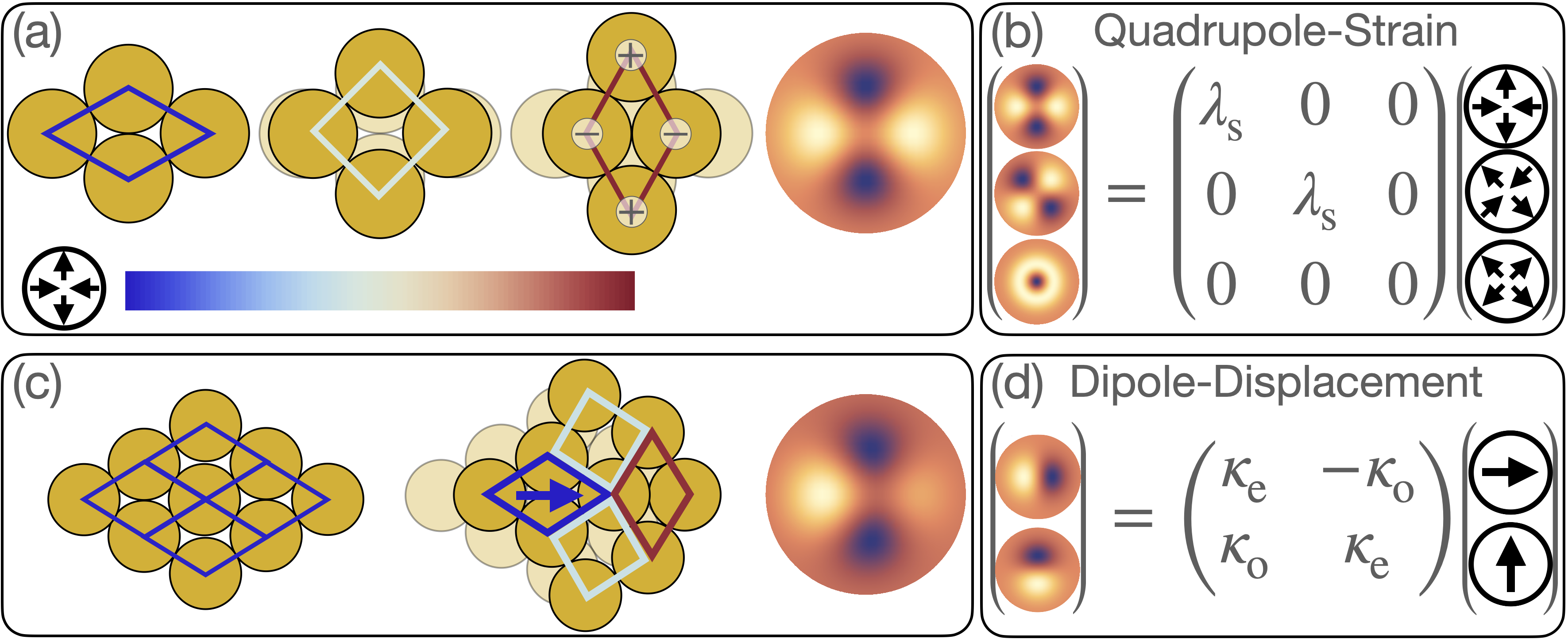}
    \caption{{The basic mechanical screening modes, their realizations, and the corresponding geometric charges. (a) Strain-induced particles rearrangement and the resultant quadrupole of angle deficit/excess marked by $\pm$. The quadrupole magnitude is proportional to the inducing strain, as indicated by the color-bar. On the right, a heat map displays induced charge, with dark/bright colors representing negative/positive charges. (b) A visual representation of the linear relation between the inducing strain modes and the induced quadrupole charges.
    (c) Displacement-induced particles rearrangement results in non-uniformly distributed quadrupoles. On average, deformation description necessitates a combination of a quadrupole and a dipole, as shown in the asymmetric charges heat map on the right. (d) A visual representation of the linear relation between the inducing displacement and the induced dipole charge $\mathbf{P} = \Gamma \mathrm{\mathbf{d}}$. In the example shown in (c) the displacement and dipole are parallel, corresponding to $\kappa_o = 0$.}}
    \label{fig:ConstRel}
\end{figure*}

We first introduce a first-principles-based theory of \textit{odd dipole screening} (ODS), for passive systems that possess internal degrees of freedom of non-conservative strain relaxation.
\MM{The elastic content of ODS assumes a reference state that can be modified in response to mechanical perturbations via particles rearrangements. The emergent elastic stress may be fully screened for certain mechanical perturbations, or partially screened for others. 
The screening mechanism is encoded in a constitutive relation that relates mechanical deformations with particles rearrangement. This relation contains an odd component $\kappa_o$ next to the usual even screening modulus $\kappa_e$.}
We show that $\kappa_e$ quantifies strain relaxation as recently proposed in \cite{livne2023geometric}, and the odd modulus $\kappa_o$ quantifies the amount of work that can be harnessed or stored within a closed deformation cycle, as attributed to \MM{particles rearrangements}. 
We quantitatively test the theory against unexplained features in the continuum strain fields in an experimental model passive amorphous material \cite{wang2020sheared}. As we will see, ODS quantitatively predicts an experimentally observed anomalous coupling between orthogonal components of the displacement field.
Importantly, the values of the new moduli display physically meaningful behavior. At small strains, $\kappa_e$ and $\kappa_o$ emerge as well-defined constants that characterize the material, as expected from a linear continuum theory. As we extend our analysis to finite strains, we find that the moduli adhere to a universal, strain-dependent relationship. Remarkably, this relationship remains independent of initial conditions or packing fraction. This revelation provides new impetus to the search for the underpinnings of plasticity in granular matter, and offers many tantalizing perspectives in a wider application of ODS to the broader field of disordered materials. Most importantly, we confirm that continuum theories of matter can describe large deformations in disordered materials experiencing particles rearrangements.

{\section{Mechanical Screening} }
One ingredient of the theory is the assumption that the primary function of localized particles rearrangements is to suppress {(or screen)} elastic fields by modifying the rest state relative to which strains are measured \cite{livne2023geometric}. A mechanical screening theory based on this concept has already been successful in describing certain types of anomalous deformation modes in amorphous systems \cite{ 22MMPRSZ,22KMPS}. The natural relaxing degrees of freedom have been shown to be quadrupolar geometric charges, which unify the description of force-dipoles, Eshelby inclusions, holes, and more \cite{moshe2015geometry, moshe2015elastic, bar2020geometric}.
In the current context of disordered matter these charges, denoted $Q^{\alpha\beta}$, describe particles rearrangements and are induced by an imposed strain as illustrated in Fig.~\ref{fig:ConstRel}(a). 
\MM{The elastic strain is then related with the strain $\uel = u - q$ where $u$ is derived from the displacement fields, and $q$ is the adjugate of $Q$. The elastic energy is quadratic in $\uel$, with the mechanical tensor $\A$ encoding bare mechanical moduli $\lambda_0,\mu_0$. Without screening these constants coincides with Lame coefficients. See  \textit{methods} for the relation between $\lambda_0,\mu_0$ and the actual observed mechanical moduli. }
A typical particles rearrangement is shown to induce a quadrupole of angle excess or deficit, thus acting as a source of stress when implemented in an elastic solid \cite{moshe2015elastic,moshe2014isf}. 
\MM{For quasistatic deformations the dissipative equations of motion reduces back to the equilibrium equations, which include a screening relation and a force-balance equation (see \textit{methods}).}
The linear relation between the inducing strain and the induced charge, illustrated in Fig.~\ref{fig:ConstRel}(b), 
establishes the basis for a theory of quadrupoles mediated mechanical screening. This relation accompanies the force balance equation, which in the presence of screening dresses the stress field and reads $\partial_{\alpha} \sigmael^{\alpha\beta} = \partial_{\alpha} (\sigma^{\alpha\beta} +\tfrac{\mu_0}{4} Q^{\alpha\beta}) = 0$. 
Our approach to mechanical screening is strongly inspired by electrostatic screening, and the regime of quadrupole screening is analogous to electrostatic dipole screening as in dielectric materials.

\begin{figure}
    \centering
    \includegraphics[width=\linewidth]{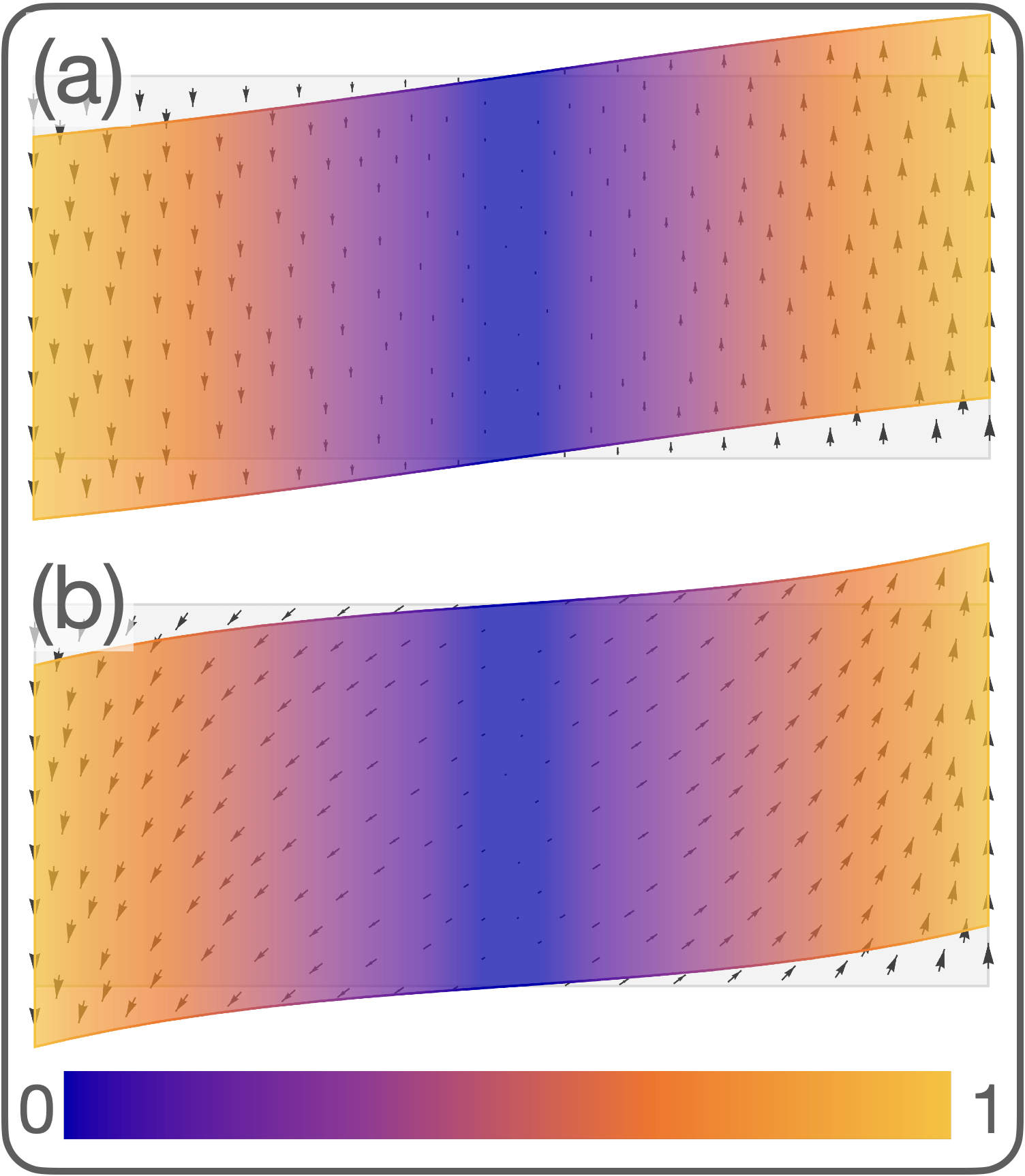}
    \caption{The displacement induced by a body shear as obtained by solving \eqref{eq:anomalous}  without (a) or with (b) odd dipole screening. Arrows indicate the displacement direction and background colors indicate displacement magnitude.}
    \label{fig:defor}
\end{figure}

A key property in this class of models is the dipole field $\mathbf{P}$ which quantifies non-uniformly distributed quadrupoles ${P}^{\alpha} = \partial_\beta Q^{\alpha\beta}$. This is illustrated in Fig.~\ref{fig:ConstRel}(c) where we show that particles displacement induces non-uniform quadrupoles. Consequently, the deformation description of a collection of particles requires a combination of a quadrupole and a dipole. 
Then, the equilibrium equation reflecting force balance is accompanied by a screening constitutive relation which relates inducing displacement with induced dipoles ${P}^\alpha = \Gamma^{\alpha\beta} \mathrm{d}_\beta$, with $\Gamma$ the screening tensor, as illustrated in Fig.~\ref{fig:ConstRel}(c,d).   In homogeneous, isotropic, conservative systems the screening tensor can be expressed as $\Gamma = \kappa_e I$ with $I$ the identity. In this case the inducing displacement is parallel to the induced dipole, as in Fig.~\ref{fig:ConstRel}(c).

In states exhibiting elastic-like behavior, for example disordered matter under strong confinement, quadrupoles will form only sparsely. Consequently, no dipoles will form, corresponding to $\kappa_e = 0$. Conversely, in situations where the cost associated with quadrupole formation is negligible, as in disordered matter with lenient confining conditions, quadrupoles form abundantly, corresponding to a finite $\kappa_e$. In this case the screening relation modifies the form of the force balance equation.
Upon integrating out the screening degrees of freedom, the force balance equation in $2D$ reduces to
\begin{eqnarray}
    	\Delta \mathbf{d} + r \nabla \left(\nabla\cdot \mathbf{d}\right) = -{\Gamma} \mathbf{d}\;.
	\label{eq:anomalous}
\end{eqnarray}
with  $r_\mathrm{2D} = \tfrac{1+\nu}{1-\nu}$ and $\nu$ the Poisson's ratio; this can be generalized to 3D \cite{22CMP}. Equation (\ref{eq:anomalous}) is a mechanical analogue of Helmholtz equation in Debye-H{\"u}ckel theory of ionic liquids.  
{An immediate and important implication of this theory is the protocol-dependent value of effective elastic moduli. In \textit{methods} we show that pure shear does not support stress whereas simple shear does.} 

\section{Odd dipole Screening}
Eq.~\eqref{eq:anomalous} is derived from a potential energy, which imposes a symmetry condition on the screening tensor $\Gamma^{\alpha\beta} = \Gamma^{\beta\alpha}$.
{However, in realistic (yield stress) materials, energy is not conserved due to particles rearrangements \cite{sun2015granular}. } During quasi-static shear, particles rearrange, and upon completing a closed loop in strain space, e.g. by sequentially applying different shear strain modes, it is likely that the system will not end up in the same state. 
This basic observation from many amorphous materials reflects the existence of multiple meta-stable states, and even fractal-like energy landscape \cite{14CKPUZ}. Consequently, during this work cycle net work can be extracted or stored in the system.

This suggests that Eq.~\eqref{eq:anomalous} with $\Gamma = \kappa_e I$ cannot provide a complete description of disordered solids. 
Therefore, as a second ingredient of the theory, we need to remove the assumption on the existence of potential energy while retaining the successful modeling picture of mechanical screening theory. To do so, we 
allow $\Gamma$ to contain an anti-symmetric (odd) screening modulus
\begin{equation}
	\Gamma  = \begin{pmatrix}\kappa_e & -\kappa_o\\
\kappa_o & \kappa_e
\end{pmatrix} \equiv \kappa\, \mathrm{Rot}(\theta_\kappa)
\end{equation}
Here  $\kappa = \sqrt{\kappa_e^2 + \kappa_o^2}$ quantifies the screening magnitude, and $\mathrm{Rot}(\theta_\kappa)$ is a rotation operator with $\tan \theta_\kappa = \kappa_o/\kappa_e$, which we term the odd screening phase. 
Geometrically, the odd screening modulus $\kappa_o$ quantifies the rotation angle between an inducing displacement field and the induced dipole.  
{From a mechanical perspective, $\kappa_o$ quantifies the amount of work extracted or stored during the work cycle due to particle rearrangements. Explicitly, $\Delta W = \oint f^\alpha d_\alpha \mathrm{d}^2 x  \propto \kappa_o $, see \emph{methods} for detailed calculation. }

\begin{figure*}
    \centering
    \includegraphics[width=\linewidth]{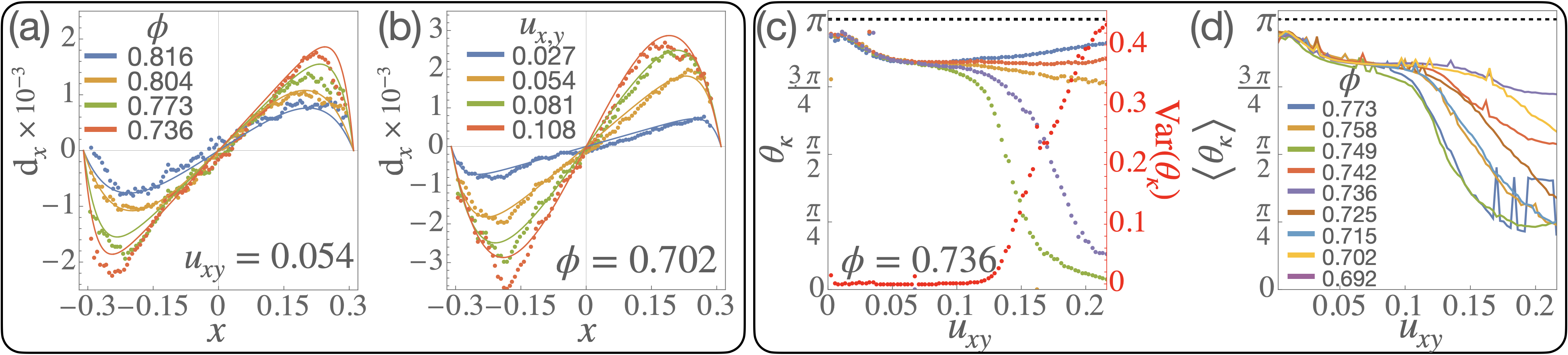}
    \caption{{\bf Odd-displacement coupling in disordered granular matter} Left panel: Theory (solid lines) and experimental data (dots) for anomalous displacement $d_x(x)$ induced by an affine shear $d_y = u_{xy}\,x$. The displacement is averaged along the $y$ direction and fitted to Eq.~\eqref{eq:sol} with respect to $\kappa$ and $\theta_\kappa$. In (a), we present the experimental and theoretical displacement fields for various packing fractions at a fixed imposed shear, and in (b), we display them for various shear loadings at a fixed packing fraction. Right panel: $\theta_\kappa$ as function imposed shear $u_{xy}$. In (c), we illustrate $\theta_\kappa$ for various initial realizations with a fixed packing fraction. The variance is shown in red, with virtually no variance observed up to strains of 10\%. In (d) we depict the ensemble-averaged screening phase $\left<\theta_\kappa\right>$ for a range of initial packing fractions, highlighting its independence not only from initial conditions but also from the initial packing fraction.}
    \label{fig:disp}
\end{figure*}

\section{ Anomalous displacement coupling} An immediate implication of ODS is the new coupling between the different displacement components in Eq.~\eqref{eq:anomalous}. 
This anomalous coupling is absent from other theories of matter in homogeneous and isotropic solids, and is crucial for describing disordered solids. 
To illustrate its implications we solve Eq.~\eqref{eq:anomalous} for the two most studied loading protocols of shearing a material either in an annular or in a rectangular domain, as studied in an immense volume of works, see for example \cite{96JNB,de1999granular, alexander1998amorphous, howell1999stress,veje1999kinematics} and references therein.
Our predictions for the case of a rectangular domain are given in detail and compared to experimental measurements; the case of an annular domain is presented in the SI. 

For a rectangular domain subjected to a body shear, as studies in \cite{18DJDZB}, the equilibrium equation \eqref{eq:anomalous} is supplemented with the conditions $d_y(x) = u_{xy} x$ and $d_x(\pm L/2) = 0$. The deformation protocols and the corresponding solutions are illustrated in Fig.~\ref{fig:defor}. In the absence of ODS $\kappa_o = 0$, the response is purely affine $d_y(x) = u_{xy} x$, with $d_x(x) = 0$, as illustrated in Fig.~\ref{fig:defor}(a). In the presence of ODS, 
we find a non vanishing solution
\begin{eqnarray}
    \mathrm{d}_x(x) =   \tau \,  \left( x  -   \frac{\sin\left( \zeta \, x\right) }{\sin \left(\zeta\right) } \right)  \;,
    \label{eq:sol}
\end{eqnarray}
with $\tau = u_{xy}\, \tan \theta_\kappa$, $\zeta=\sqrt{ \cos \theta_\kappa \,\kappa/(1+r) } \,  L/2$, and $x$, and $d$ are measured in units of $L/2$. This solution is illustrated in Fig.~\ref{fig:defor}(b).

\section{Results}
The anomalous displacement coupling predicted by ODS align with observations from recent experiments on two-dimensional granular model system subjected to simple body shear \cite{wang2020sheared}. In this experiment
nonzero displacement fields $d_x \neq 0$ emerged, and thus serves as the ultimate test for the theory of ODS. 
We investigate a disordered system of 2D \MM{pre-jammed} granular matter composed of bi-dispersed frictional disks. The body shear protocol is applied quasistatically by uniformly deforming the underlying table (see \emph{method} for further details). We explore a range of imposed strains, up to 22\%, and initial packing fractions within the range of $0.692 < \phi < 0.816$. To assess sensitivity to initial conditions, each experiment is repeated five times.
In the experiment, we measure the displacement field of each particle and, due to the system's symmetry, we average it along the $y$ direction before comparing it with the prediction of \eqref{eq:sol}. For each measurement of the displacement field, we fit the parameters $\eta$ and $\tau$, from which we extract the screening moduli $\tilde{\kappa} = \kappa/(1+r)$ and $\theta_\kappa$. \MM{To measure $r$ further independent perturbations are required, e.g. isotropic compression, but are beyond the scope of our experiment. The observation that the system is unjammed aligns with the theory, as pure dipole screening predicts a vanishing observable resistance to pure-shear deformation (see \textit{methods}).} 
In left panel of Fig.~\ref{fig:disp}, we present plots of the induced transverse displacement $d_x$ as a function of position $x$ together with our predictions. In (a) we present plots for various values of packing fractions with a fixed imposed strain, and in (b) we present plots for various values of imposed strains with a fixed packing fraction. We observe that \eqref{eq:sol} fits very well and perfectly reproduces the functional form of the anomalously induced displacement.

The success in recovering the functional form of the displacement field by fitting the screening moduli is a promising step. To further demonstrate that that the screening moduli are proper macroscopic quantifiers for the state of jammed granular matter, in the \emph{method} section, we show that for small strains, both screening moduli remain independent of the initial conditions and assume constant values as a function of strain, independent of the initial configurations.
Surprisingly, our analysis reveals that these explicit moduli are also unaffected by variations in the initial packing fraction, with values of $\eta \approx 11 L^2$ and $\theta_\kappa \approx 0.94  \pi$. 
Notably, the value of the odd screening phase $\theta_\kappa$ indicates that the screening effect is predominantly even $\kappa_o/\kappa_e\approx -0.2$.

Another surprising discovery is that the theory, initially designed for small deformations based on linearized strain-stress and dipole-displacement relations, describes the displacement field accurately even at substantial strains exceeding 20\%.
Drawing an analogy to non-linear optics, where the permittivity can explicitly depend on the field's magnitude, it is plausible that at large deformations, the screening moduli might exhibit a dependence on the strain.
As expected, we observe that at finite strains $\theta_\kappa$ and $\kappa$ deviate from their constant values at low strains (see \emph{methods} for plots of $\kappa$). In Fig.~\ref{fig:disp}(c) we show plots of $\theta_\kappa$ for five different realizations of identical initial packing fraction $\phi = 0.736$.
Remarkably, we find that in all experiments the dependence of $\theta_\kappa$ on the imposed strain $u_{xy}$ is identical, with a variance that is practically zero, up to strain of $~10\%$, \MM{which is possibly the onset of fragility}.  At larger strains $\theta_\kappa(u_{xy})$  still acquire a clear functional form, though with different functional form for different initial conditions. 
This observation suggests that the system is self averaging up to strains as high as $10\%$, thus allowing its description within a continuum theory even for large deformations. The origin of the precise functional form of $\theta_\kappa(u_{xy})$ is an exciting avenue for much future work, while the large strain effectiveness of the continuum modeling directly reduces the need for expensive mesh-free methods to predict material behavior at large strains. 
The situations becomes even more surprising when $\theta_\kappa(u_{xy})$ is computed for different initial packing fraction, where we discover that all initial conditions and all packing fractions have exactly the same form, as shown in Figure \ref{fig:disp}(d). This universal form indicates a universal underlying nonlinear screening mechanism independent of the packing fraction.

The weak decay of the screening phase $\theta_\kappa$, transitioning from approximately $\pi$ to $0.8\pi$, signifies an important new physics of screening within disordered media. Screening can be equivalently achieved by either generating new charges, depending on the protocol, or releasing pre-existing ones. In the latter scenario, the availability of pre-existing charges is limited, resulting in an expected weakening of the screening effect during the process.
Consequently, the decline in $\theta_\kappa$ signifies that screening predominantly occurs through the release of pre-existing charges, which were trapped within the system during the jamming process. This conclusion finds independent support in the relationship between the screening magnitude, $\kappa$, and the applied shear stress, $u_{xy}$, which exhibits an exponential decay, indicating the lack of existing charges to be released (see App.\ref{secA1}).
Our theory, therefore, posits that a fundamental formulation of screening theory, rooted in microscopic principles, necessitates a comprehensive understanding of the initial jammed state's pre-existing charges. It is noteworthy that the charges induced as a response to shear jamming and isotropic pressure may fundamentally differ, thus prompting the need for a quantitative investigation into the connection between jamming and induced charges.

\section{Summary} In conclusion, our ODS theory correctly captures small and large strain behavior of a model disordered media via the introduction of only one new modulus. The new screening modulus has a clear physical microscopic interpretation and is observed to be indeed universal for the experimental model system used. Our work calls for much future work, such as applying to the wide range of other disordered materials like colloidal glasses, foams, emulsions and microgels. ODS can explain unusual deformation modes that can be captured by performing new or different types of rheological experiments. It is clearly possible expand our findings by developing a nonlinear version of ODS to comprehensively analyze the failure modes of disordered solids. As our theory now only works in quasi-static deformation, there is an obvious perspective to incorporate dynamical and dissipative mechanisms. Crucially, since odd-elasticity and odd-screening are two independent extensions of two different theories, they should be integrated and applied to experimental active solids that include relaxational modes, such as epithelial tissue.

\section{Methods}\label{sec11}
\begin{figure}[h]
    \centering
    \includegraphics[width=\linewidth]{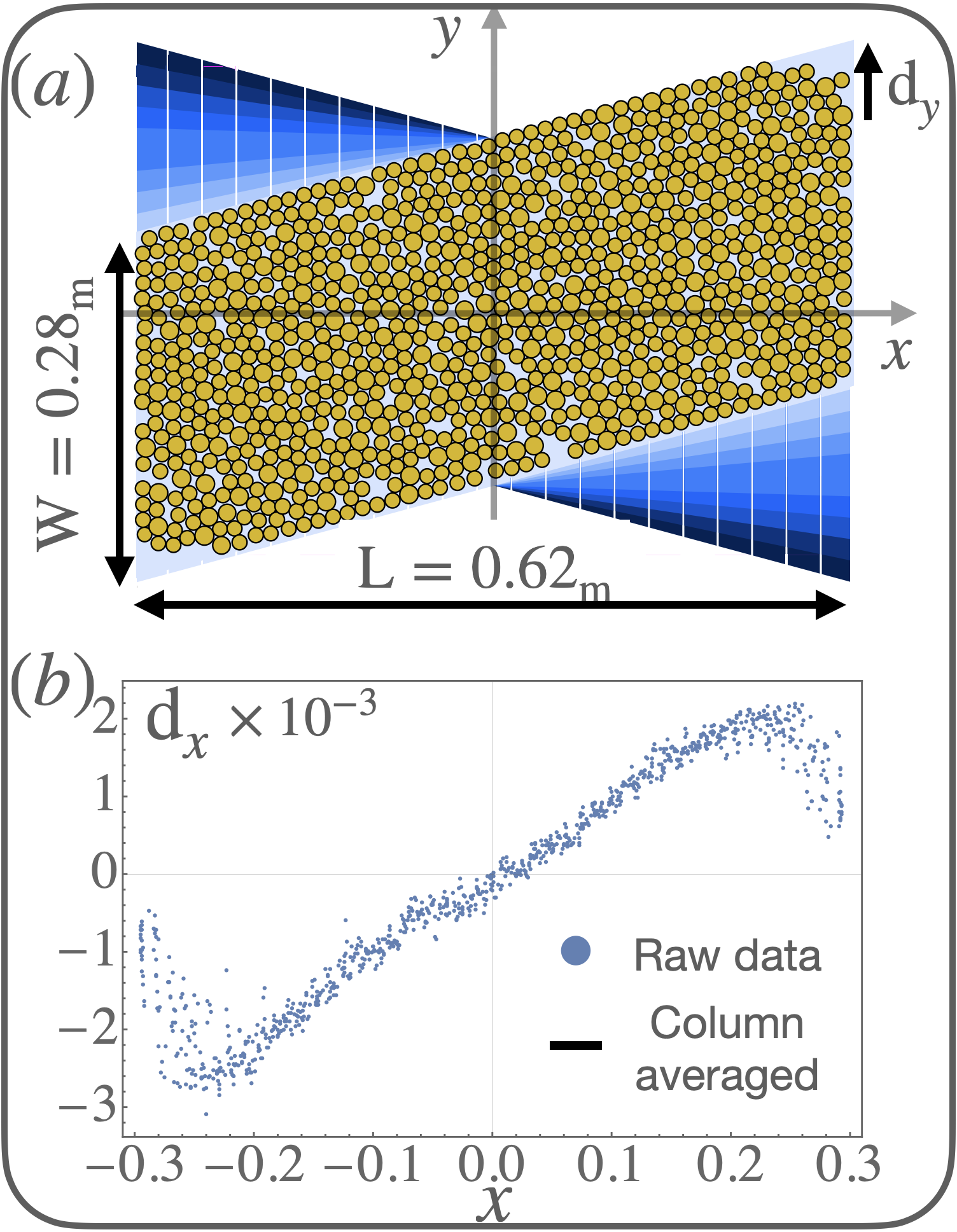}
    \caption{(a) Schematic of special apparatus for applying simple shear strain to a collection of photoelastic discs at the different strain steps (strain steps increase from light to dark colors). In the sketch, the slats are drawn much larger relative to the particles than in the real experiment. The $x-y$ axes indicate the coordinate system in the lab frame, where simple shear is applied along the y axis. (b) Example of particle displacements $d_x$ in a system with initial packing fraction $\phi = 0.742$, after a shear strain of $5\%$.}
    \label{fig:experiment}
\end{figure}

\textbf{Experiment Setup.} 
We applied quasi-static simple shear to a bi-dispersed granular system composed of elastic discs. Shear was performed in an apparatus that suppresses shear bands allowing us to study large strains before the onset of shear bands ~\cite{ren13prl}. Particles were carried by separate slats that form the shear box base. These slats moved affinely in accordance with the applied shear~\cite{ren13prl, wang20prl}. Discs were cut from elastic sheets (Vishay PSM-4), resulting in a friction coefficient $\mu \approx 0.7$ \cite{ren13prl, zadeh19_gm}. The system contained approximately $45\times20$ discs with a diameter ratio $1:1.25$ and a large to small number ratio $1:3.3$ to prevent crystallization. Every run at a given packing fraction $\phi$ was repeated five times, with the initial stress-free state being prepared anew for each run. Shear was applied quasi-statically in the $y$ direction to a shear box (Fig.~\ref{fig:experiment}(a)). Starting from a stress-free state, the system was sheared by a strain step of $\delta\gamma = 0.0027$. Then the system was left to relax for six seconds, followed by taking an images which reveal information on particle position. Such a process -- stepwise shearing, relaxing and imaging -- was repeated until a certain amount of total strain was achieved. The maximum shear strain $\gamma$ that studied in this work is $0.24$ (24\%). From the images, we tracked particle positions and computed the displacement field. All tracking data are available upon request. \cite{howell97pg,howell99prl,zadeh19_gm}.

\textbf{(Even) Mechanical Screening}
This section is supplemented with a \emph{Mathematica} notebook available online, where all calculations, including lengthy ones, are given in full detail.   

The deformation of elastic solid is described by a displacement field $\mathbf{d}$. The elastic strain, which measures deviations of actual lengths from their rest values, is $u_{\alpha\beta} = \tfrac{1}{2} (\partial_\alpha d_\beta + \partial_\beta d_\alpha)$.
The elastic energy density is then
\begin{eqnarray}   
    \mathcal{W} =  \frac{1}{2}\A^{\alpha\beta\gamma\delta} u_{\alpha\beta} u_{\gamma\delta}  \;.
\end{eqnarray}
where $\A$ is the elastic tensor.

Mechanical screening is the process at which internal degrees of freedom are available to relief strains. Then one should distinguish between the configurational strain $u$ derived from the displacement field, and the elastic strain $\uel$ for which elastic energy penalizes for
\begin{eqnarray}
    \uel = u - q
\end{eqnarray}
Here $q$ is the local anelastic (screening) strain. In general, it has three independent degrees of freedom, corresponding to relaxing isotropic strain and the two modes of shear strain. While the theory is applicable to arbitrary form of $q$, particles rearrangements allow to relax only shear strain and not isotropic one, thus we assume $\mathrm{trace}(q) = 0$.

The elastic energy is 
\begin{eqnarray}   
    \mathcal{W}_\mathrm{el} =  \frac{1}{2}\A^{\alpha\beta\gamma\delta} \uel_{\alpha\beta} \uel_{\gamma\delta} \;.
\end{eqnarray}
    
The induced local anelastic strain $q$ creates elastic fields that relaxes the total strain but cost energy due to its nucleation. Therefore the total energy contain another term for the nucleation cost associated with the induced strain relaxation. For that we note that a distribution of relaxing $q$'s induces elastic strains whose sources (geometric charges) are $Q^{\alpha\beta} = \varepsilon^{\alpha\mu}\varepsilon^{\beta\nu} q_{\mu\nu}$ \MM{(that is $q$ is the adjugate of Q: $q = \mathrm{adj} \,Q$)} \cite{moshe2015elastic}. If the $Q$ field is non uniform, according to the multipole expansion of geometric charges a hierarchy of multipoles will form with an effective dipole field $P^\alpha = \partial_\beta Q^{\alpha\beta}$ and an effective monopole field $M = \partial_{\alpha\beta} Q^{\alpha\beta}$ \cite{moshe2015elastic}.
The nucleation cost is therefore of the form
\begin{equation}
    \mathcal{F} = \frac{1}{2} \Lambda_{\alpha\beta\gamma\delta} Q^{\alpha\beta}Q^{\gamma\delta} + \frac{1}{2} \tilde\Gamma_{\alpha\beta} P^{\alpha} P^{\beta} + \frac{1}{2} \Upsilon M^2 \;.
    \label{eq:F}
\end{equation}
Since no isotropic quadrupole can form, the most general forms of $\Lambda$ and $\Gamma$ in isotropic and homogeneous media are
\begin{eqnarray}
    \Lambda_{\alpha\beta\gamma\delta}  &=& \frac{1}{2}  \lambda \left(\delta_{\alpha\gamma} \delta_{\beta\delta} + \delta_{\alpha\delta} \delta_{\beta\gamma}\right) \\ 
    \tilde\Gamma_{\alpha\beta} &=& \tilde\kappa_e \, \delta_{\alpha\beta}
\end{eqnarray}
Monopole screening has not been observed in solid like phases and therefore is beyond the scope of the current work, i.e. $\Upsilon = 0$. 
\MM{In this work we are interested in a dissipative system that experience plastic deformations. As such the governing equations are dissipative ones and take the form
\begin{eqnarray}
	\dot{\mathbf{d}} = - \chi_d \frac{\delta \, W}{\delta \, {\mathbf{d}}}\;, \quad \quad \dot{Q}^{\alpha\beta} = - \chi_Q \frac{\delta \, W}{\delta \, Q^{\alpha\beta}}
\end{eqnarray}
with $W = \mathcal{W}_\mathrm{el} + \mathcal{F}$. In the limit of quasi-static deformations these equation reduces back to equilibrium equations. Consequently, the emergent quadrupole field, for example, is permanent, and it will not be relaxed upon releasing the mechanical perturbations. 
}
Regardless of the specific form of \eqref{eq:F}, the 2d equilibrium equation, obtained by energy minimization with respect to the displacement is
\begin{eqnarray}
    \Delta \mathbf{d} + \frac{1+\nu}{1-\nu} \nabla \left(\nabla \cdot \mathbf{d}\right)  = -\mathbf{P}
    \label{eq:FB}
\end{eqnarray}
where $\nu$ is the Poisson's ratio.

This equation is supplemented by a constitutive screening relation obtained by variating the energy with respect to the screening degrees of freedom $Q$. 
For quadrupole screening we find
\begin{eqnarray}
    \mathbf{P} = - \frac{1}{1 + 4 \tilde{\lambda} (1+\nu)} \Delta \mathbf{d}\;,
\end{eqnarray}
with $\lambda = Y\,\tilde \lambda$.
Upon substituting in \eqref{eq:FB} we find that the form of the equation remains the same but the coefficients are renormalized by the screening. 

In the case of dipole screening, where the nucleation cost of quadrupoles is negligible compared with that of dipoles, we find
\begin{eqnarray}
    \mathbf{P} = \frac{Y}{2\tilde\kappa_e (1+\nu)} (\mathbf{d} - \mathbf{d}_0)
\end{eqnarray}
where $\mathbf{d}_0$ is an integration constant. This relation represent the relation between an imposed displacement and an induced elastic dipole, as illustrated in Fig.~\ref{fig:ConstRel}.
We set $\mathbf{d}_0 = 0$, thus the constitutive relation loses its invariance under translations and rotations. We denote 
\begin{eqnarray}
    \mathbf{P} = \Gamma \mathbf{d}
\end{eqnarray}
with 
$\Gamma_{\alpha\beta} =\kappa_e \delta_{\alpha\beta} $ and $\kappa_e = \frac{Y}{2\tilde\kappa_e (1+\nu)}$.
Upon substituting in \eqref{eq:FB} we find
\begin{eqnarray}
    \Delta \mathbf{d} + \frac{1+\nu}{1-\nu} \nabla \left(\nabla \cdot \mathbf{d}\right)  = -\Gamma \mathbf{d}
    \label{eq:Anomalous}
\end{eqnarray}
that is we recovered \eqref{eq:anomalous}.

\MM{\textbf{Measurement of mechanical and screening moduli}}
\MM{In classical elasticity, the components of the elastic tensor are associated with the resistance of the system to specific mechanical perturbations. For example, in the absence of screening, the resistance to pure shear and uniaxial strain are obtained by substituting in the energy the corresponding strains
\begin{equation}
	u_\mathrm{PS}  = \begin{pmatrix}u_0 & 0\\
		0& -u_0
	\end{pmatrix}\;,  \quad \quad u_\mathrm{UNI}  = \begin{pmatrix}u_0 & 0\\
	0& 0
\end{pmatrix} 
\end{equation}
The corresponding expressions  for the energy are
\begin{equation}
	W_\mathrm{PS} =\frac{1}{2} 2\mu_0 u_0^2 \;, \quad \quad W_\mathrm{UNI} =\frac{1}{2} (\mu_0 + \lambda_0) u_0^2  \;.
\end{equation}
where $\mu_0,\lambda_0$ are the shear and bulk moduli encoded in the elastic tensor $\A$.
It is widely accepted that if the observed resistance to pure-shear deformation vanishes, the corresponding modulus $\mu_0$ is zero, and similarly with $\lambda_0$. 
When it comes to disordered solids, it is known that the same system may support one mode of deformation and not others, implying for an ill-defined notion of the elastic moduli. Nevertheless, the automatic association between measurement and components of the elastic tensor are still accepted.
Here we show that this misconception is resolved by the theory of mechanical screening. 
 In the presence of quadrupole and dipole screening, that is $\Lambda, \tilde{\Gamma} \neq  0$   in \eqref{eq:F}, upon imposing the uniform deformations above, one has to minimize with respect to the screening quadrupole field the total energy
\begin{eqnarray}
	W &=& \mathcal{W}_\mathrm{el} + \mathcal{F}  = \frac{1}{2} \A^{\alpha\beta\gamma\delta} (u_{\alpha\beta} - q_{\alpha\beta}) (u_{\gamma\delta} - q_{\gamma\delta}) \notag\\ &+&  \frac{1}{2} \Lambda_{\alpha\beta\gamma\delta} Q^{\alpha\beta} Q^{\gamma\delta}  +  \frac{1}{2} \tilde\Gamma_{\alpha\beta} P^{\alpha} P^{\beta}\;.
\end{eqnarray}
In this case the resistance to the same imposed strains is screened by uniformly distributed quadrupoles. The energy in each case is
\begin{equation}
	W_\mathrm{PS}  =\frac{1}{2} \frac{2 \mu_0 \lambda}{\mu_0  + \lambda} u_0^2  \;, \quad \quad W_\mathrm{UNI}=\frac{1}{2}\left( \lambda_0 + \mu_0 \frac{2  \lambda + \mu_0}{2(\lambda + \mu_0)} \right)u_0^2 \;.
\end{equation}
Within this picture, the vanishing resistance to pure shear implies that either $\mu_0=0$, or $\lambda=0$. We claim that what is commonly accepted as an unjammed state in granular and disordered matter is possible with $\lambda=0$, that is pure quadrupole screening, with finite bare mechanical  moduli $\mu_0,\lambda_0$.
To measure all coefficients $\mu_0, \lambda_0, \lambda$, an additional measurement is required, e.g. isotropic compression. In that case screening is inactive, and we find $W_\mathrm{ISO} = \tfrac{1}{2} (2\mu_0 + 4\lambda_0) u_0^2$. \\
The measurement of $\tilde{\Lambda}$ in the presence of dipole screening cannot rely on uniform deformations, and require situations where the induced quadrupoles have non-vanishing divergence. The protocols studied in Appendix B provide the predicted deformation fields for different geometries and perturbations. Upon successfully fitting the observed and predicted displacement fields  one can extract the components of $\tilde{\Gamma}$, as done in the \textit{Results} section in the main text.}

\textbf{Work cycles and odd screening}
When a material is deformed, the power exerted
by the elastic forces is
\begin{equation}
\dot{W}=\intop f^{\alpha}\dot{d}_{\alpha}dS
\end{equation}
where $f$ the force is given by 
\begin{equation}
f^{\alpha}=\partial_{\beta}\sigmael^{\alpha\beta}
\end{equation}
with the elastic stress $\sigmael = \A \uel$, satisfying  $\partial_\alpha\sigmael^{\alpha\beta} = 0$, which is equivalent to \eqref{eq:FB}. When expressed in terms of $\sigma = \A u$ we find $\partial_\beta\sigma^{\alpha\beta}  + \tfrac{Y}{4(1+\nu)} \mathbf{P} = 0$.
With these notations we rewrite the power exerted by the elastic forces
\begin{equation}
\dot{W}=\intop  \left(\partial_\beta\sigma^{\alpha\beta}  + \tfrac{Y}{4(1+\nu)} {P}^\alpha\right) \dot{d}_{\alpha}dS
\end{equation}
we further simplify this expression by expressing $\sigma$ and $\mathbf{P}$ in terms of the displacement field 
\begin{equation}
\dot{W}=\intop  \left(\A^{\alpha\beta\gamma\delta} \partial_\beta \partial_{\gamma}d_{\delta}  + \tfrac{Y}{4(1+\nu)} \Gamma^{\alpha\beta}{d}_\beta\right) \dot{d}_{\alpha}dS
\end{equation}
Upon integrating by parts the first term  we find
\begin{equation}
\dot{W}=\intop  \left(-\A^{\alpha\beta\gamma\delta}  \partial_{\gamma}d_{\delta} \partial_\beta\dot{d}_{\alpha} + \tfrac{Y}{4(1+\nu)} \Gamma^{\alpha\beta}{d}_\beta \dot{d}_{\alpha}\right) dS
\label{eq:power}
\end{equation}
Here we assumed displacement-controlled boundary conditions, thus the boundary term vanishes. 
Next we calculate the total elastic work done during a closed deformation. For the first term
\begin{eqnarray}
\Delta W_1 &=& \oint \dif t\, \dot{W}_1  = -\oint \dif t\,\intop  \A^{\alpha\beta\gamma\delta}  \partial_{\gamma}d_{\delta} \partial_\beta\dot{d}_{\alpha} dS \notag\\
&=& \oint \dif t\,\intop  \A^{\alpha\beta\gamma\delta}  \partial_{\gamma}\dot{d}_{\delta} \partial_\beta {d}_{\alpha} dS \notag\\
&=& \oint \dif t\,\intop  \A^{\gamma\delta\alpha\beta} \partial_\beta \dot{d}_{\alpha}  \partial_{\gamma}{d}_{\delta}  dS
\end{eqnarray}
The third equality holds from (time) integration by parts, and the fourth equality obtained by relabeling the indexes. 
We deduce that
\begin{equation}
\Delta W_1 =  -\oint \dif t\,\intop  \frac{1}{2}\left(\A^{\alpha\beta\gamma\delta} - \A^{\gamma\delta\alpha\beta}\right)  \partial_{\gamma}d_{\delta} \partial_\beta\dot{d}_{\alpha} dS 
\end{equation}
From the major symmetry of the elastic tensor this term is identically zero. 

The second term in \eqref{eq:power} is straightforward to handle. The total work is
\begin{eqnarray}
\Delta {W}_2 &=& \oint \dif t\, \dot{W}_2  = \oint \dif t\, \intop \tfrac{Y}{4(1+\nu)} \Gamma^{\alpha\beta}{d}_\beta \dot{d}_{\alpha} dS
\end{eqnarray}
Integrating by parts the time-integral and relabeling the indexes we find
\begin{eqnarray}
\Delta {W}_2 = - \oint \dif t\, \intop \tfrac{Y}{4(1+\nu)} \Gamma^{\beta\alpha}{d}_\beta \dot{d}_{\alpha} dS
\end{eqnarray}
Hence
\begin{eqnarray}
\Delta {W}_2 &=&  \oint \dif t\, \intop \tfrac{Y}{8(1+\nu)} \left(\Gamma^{\alpha\beta} - \Gamma^{\beta\alpha}\right){d}_\beta \dot{d}_{\alpha} dS \notag\\&=& \oint \dif t\, \intop \tfrac{Y}{4(1+\nu)} \Gamma_\mathrm{o}^{\alpha\beta}{d}_\beta \dot{d}_{\alpha} dS
\end{eqnarray}
where $\Gamma_\mathrm{o}$ is the odd part of $\Gamma$. In our work this integral is proportional to $\kappa_o$.

{{\bf Acknowledgments}
	This work was supported by Israel Science Foundation Grant No. 1441/19. MM thank Gil Cohen for stimulating discussions.}

\appendix

\section{Values of screening moduli at small and large strains}\label{secA1}
The screening moduli in our theory, as in any other linear theory, hold meaning only when they remain constant. However, when dealing with large strains, there is no assurance that constant (spatially uniform) values for the elastic and screening moduli can adequately describe the intricate and non-affine deformation field. Nonetheless, we observe that the displacement field accurately adheres to our theory at large strains, albeit with strain-dependent moduli. This mirrors the behavior seen in non-linear optics, where a material's permittivity remains constant under weak fields but varies with the field magnitude under strong fields.

\begin{figure*}
    \centering
    \includegraphics[width=\linewidth]{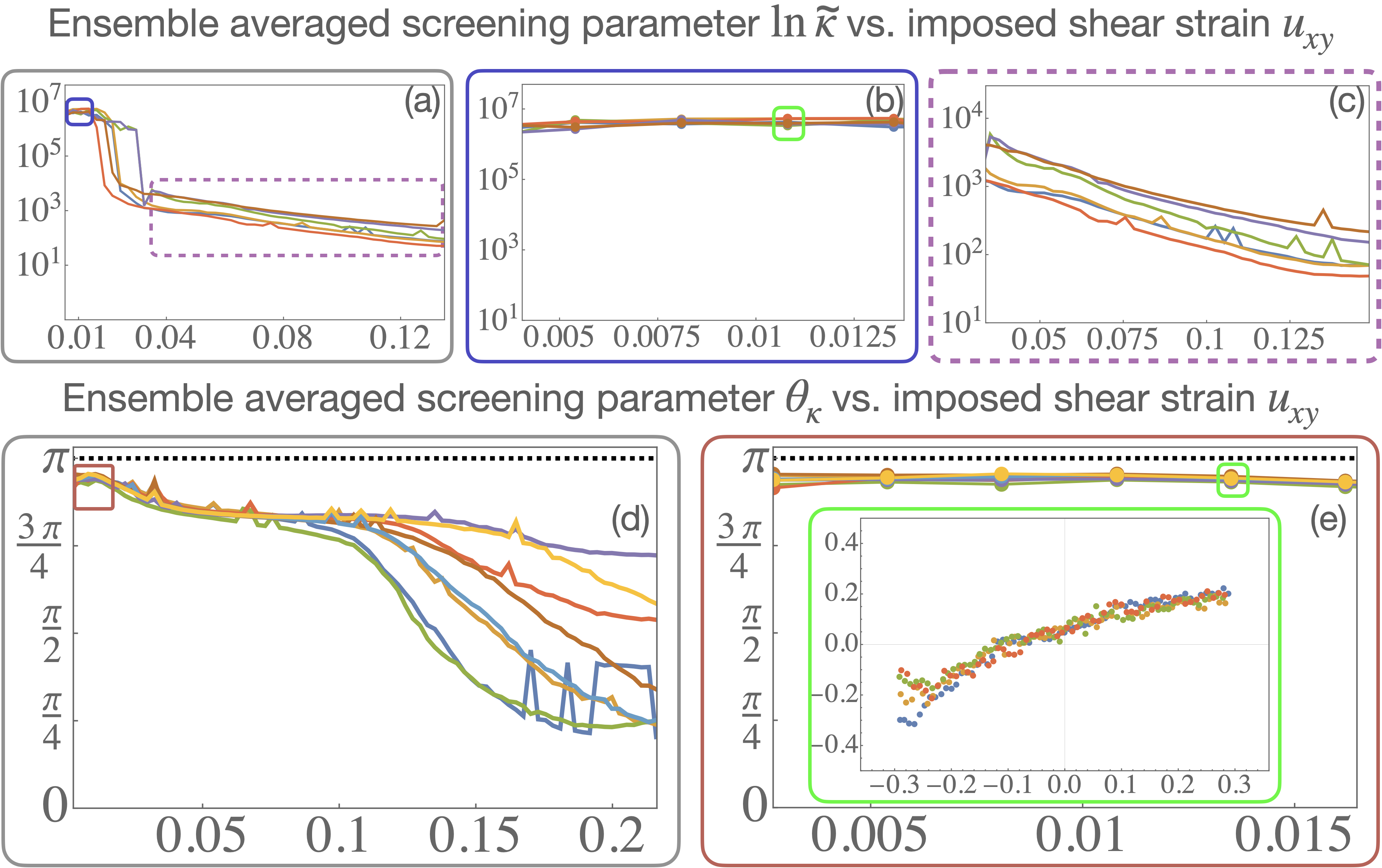}
    \caption{The ensemble averaged screening moduli as function of strains. Top panel shows the screening magnitude $\left<\tilde{\kappa}\right>$ for strain up to $\% 13$. (a) The full range, (b) zoom in on small strains regime, and (c) zoom in on large strains regime (c). Bottom panel shows the screening phase $\left<\tilde{\kappa}\right>$ for large strains (d) and small strain (e). 
    experimentally measured displacement fields $d_x(x)$ for a shear strain of $1\%$, confirming the independence of the results from the initial packing fraction.}
    \label{fig:SI-ScreeningModuli}
\end{figure*}

In Fig.~\ref{fig:SI-ScreeningModuli}(a), we depict the ensemble-averaged screening magnitude $\left<\kappa\right>$ as a function of imposed shear strain, covering strains up to 13\%, presented on a semi-logarithmic scale. Each plot corresponds to a distinct initial packing fraction.
In (b), we zoom in on the small strain region, and present $\left<\kappa\right>$ up to $1.3\%$. The constant value of the screening magnitude justifies the assumption of employing continuum mechanical screening theory for small strain analysis of the experimental system.
In (c), we zoom in on the region of large strains, revealing that the exponential decay is consistent across all initial packing fractions, albeit with different prefactors indicating potential dependence on initial conditions.

Similarly, in Fig.~\ref{fig:SI-ScreeningModuli}(d), we display the ensemble-averaged $\left<\theta_\kappa\right>$ for various initial packing fractions. In (e), we focus on the small strain region, presenting $\left<\theta_\kappa\right>$ up to $1.5\%$. Notably, we observe that this screening modulus remains constant in this range. The inset showcases experimentally measured displacement fields for a shear strain of $1\%$ enclosed by the green box, further confirming the independence of the results from the initial packing fraction.

\section{Odd dipole screening in sheared rectangular and annular domains}
\label{secA2}
In this section we solve \eqref{eq:anomalous} for rectangular and annular domain subjected to shear strain from the boundary.

For a rectangular domain we solve \eqref{eq:anomalous} with the boundary conditions
\begin{eqnarray}
    d_x(L) &=& 0\, , \quad d_x(-L) = 0\\
    d_y(L) &=& d_0\, , \quad d_y(-L) = -d_0
    \label{eq:rectBC}
\end{eqnarray}
The solution is
\begin{eqnarray}
    d_x(x) &=& \frac{d_0 \kappa_o}{\eta} \left(\frac{\sin(x/l_1)}{\sin(L/l1)} - \frac{\sin(x/l_2)}{\sin(L/l2)}\right) \\ 
    d_y(x) &=& \frac{d_0 }{2} \left(\left(1 - \frac{\kappa_e r}{\eta}\right)  \frac{\sin(x/l_1)}{\sin(L/l1)} + \left(1 + \frac{\kappa_e r}{\eta}\right)  \frac{\sin(x/l_2)}{\sin(L/l2)}\right)
\end{eqnarray}
with 
\begin{eqnarray}
    l_1^{-1} &=& \sqrt{\frac{\kappa_e (2+r)  - \eta}{2(1+r)}}\\
    l_2^{-1} &=& \sqrt{\frac{\kappa_e (2+r)  + \eta}{2(1+r)}}\\
    \eta &=& \sqrt{\kappa_e^2 r^2 - 4 \kappa_o^2 (1+r)}
\end{eqnarray}
We see that for $\kappa_o=0$ we get $l_1 = l_2$ and then $d_x=0$, that is, the anomalous coupling appears only in the presence of odd screening. 

\begin{figure*}
    \centering
    \includegraphics[width=\linewidth]{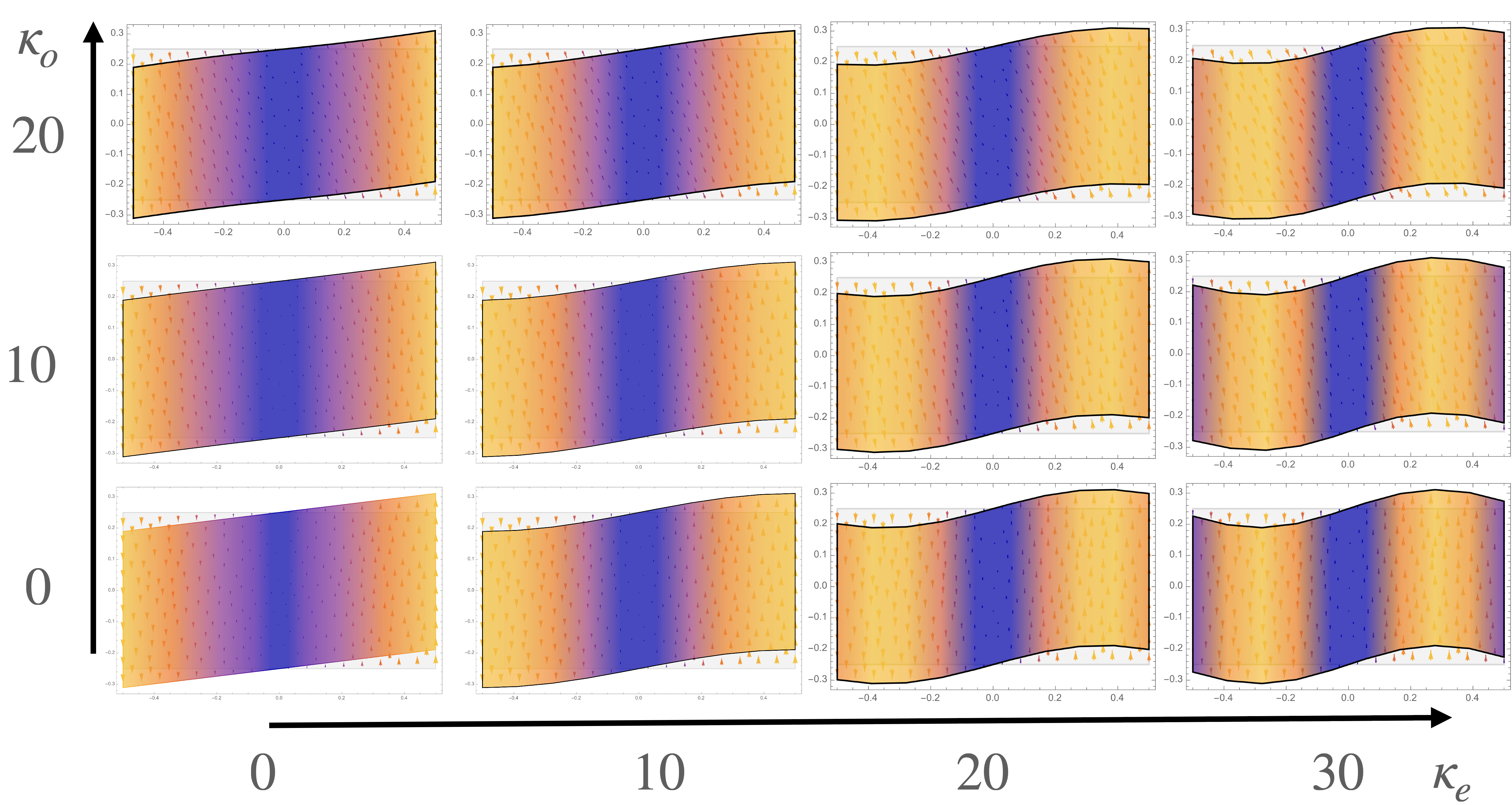}
    \caption{Solutions for \eqref{eq:anomalous} with boundary conditions \eqref{eq:rectBC}, with $d_0 = 0.05$, $r=1$, and $L=0.5$. Each block corresponds  to a solution with different values of $\kappa_e$ and $\kappa_o$.  }
    \label{fig:rectangular}
\end{figure*}

In Fig.~\ref{fig:rectangular} we plot the displacement fields for different values of the screening parameters, with $d_0 = 0.05$, $r=1$, and $L=0.5$.
For large values of $\kappa_e$ and small $\kappa_o$ we find significant deviation from the classical affine response, with non-monotonous displacement field, though with $d_x(x)=0$. For non-zero odd screening $\kappa_o\neq0$ the displacement field rotates and induces a non-zero $d_x$ component. 

\begin{figure*}
    \centering
    \includegraphics[width=0.9\linewidth]{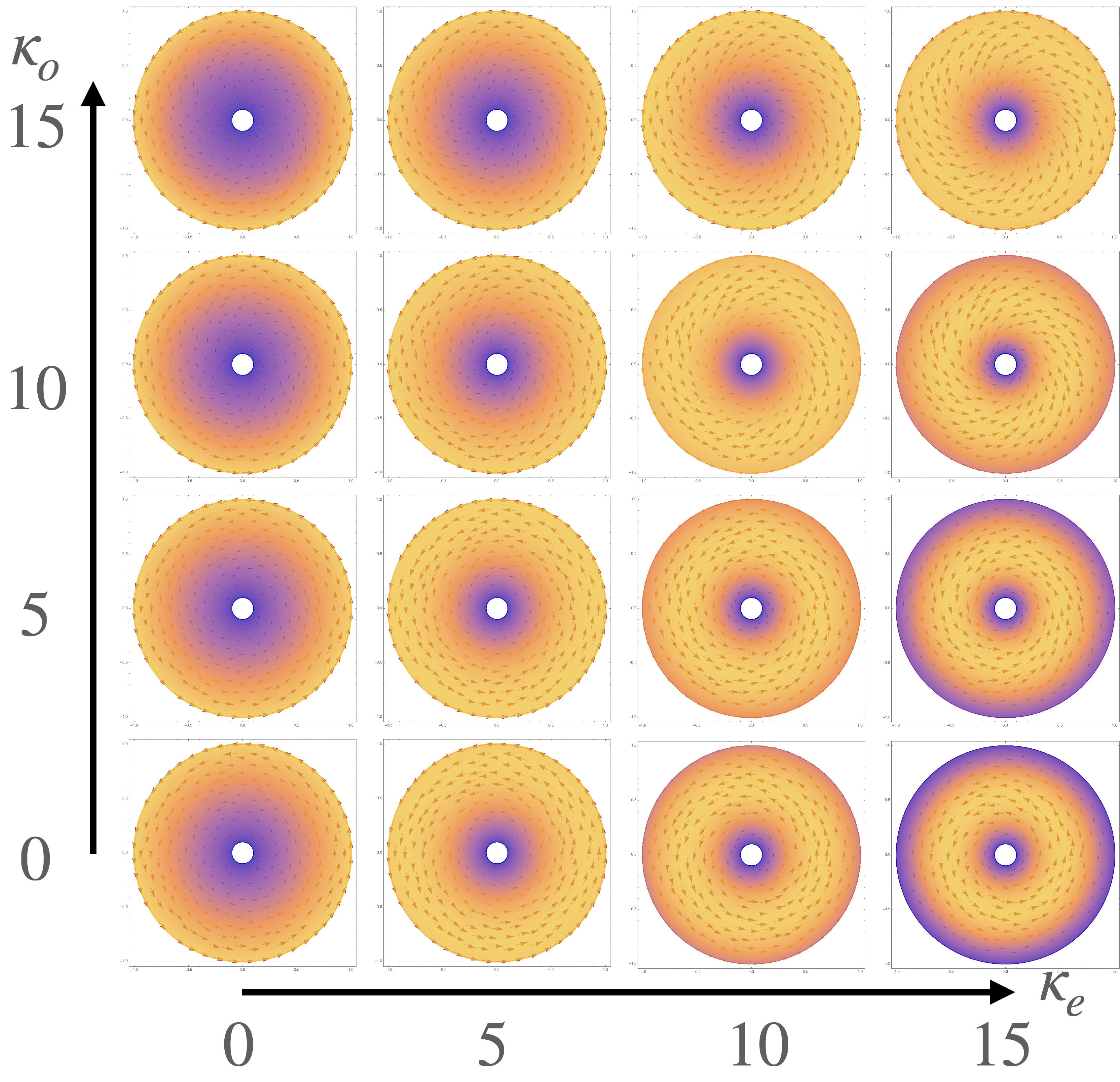}
    \caption{Solutions for \eqref{eq:anomalous} with boundary conditions \eqref{eq:annulusBC}, with $d_0 = 0.05$, $r=1$, and $\rout=1$ and $\rin = 0.1$. Each block corresponds  to a solution with different values of $\kappa_e$ and $\kappa_o$.  }
    \label{fig:annular}
\end{figure*}

For a rectangular domain we solve \eqref{eq:anomalous} with the boundary conditions
\begin{eqnarray}
    d_r(\rin) &=& 0\, , \quad d_r(\rout) = 0\\
    d_\theta(\rin) &=& d_0\, , \quad d_\theta(\rout) = 0
    \label{eq:annulusBC}
\end{eqnarray}
We solve this boundary value problem numerically, and in Fig.~\ref{fig:annular} we plot the solutions for different values of the screening parameters, with $d_0 = 0.05$, $r=1$, and $\rout=1$ and $\rin = 0.1$.
Similar to the rectangular case, at large values of $\kappa_e$ and small $\kappa_o$ we find significant deviation from the classical affine response, with non-monotonous displacement field, with $d_r(r)=0$. For non-zero odd screening $\kappa_o\neq0$ the displacement field rotates and induces a non-zero radial component $d_r$, with spiral stream lines. 


\begin{thebibliography}{10}
	
	\bibitem{90Camp}
	Charles~S Campbell.
	\newblock Rapid granular flows.
	\newblock {\em Annual Review of Fluid Mechanics}, 22(1):57--90, 1990.
	
	\bibitem{92JN}
	Heinrich~M Jaeger and Sidney~R Nagel.
	\newblock Physics of the granular state.
	\newblock {\em Science}, 255(5051):1523--1531, 1992.
	
	\bibitem{96JNB}
	Heinrich~M. Jaeger, Sidney~R. Nagel, and Robert~P. Behringer.
	\newblock Granular solids, liquids, and gases.
	\newblock {\em Rev. Mod. Phys.}, 68:1259--1273, Oct 1996.
	
	\bibitem{13AFP}
	Bruno Andreotti, Yo{\"e}l Forterre, and Olivier Pouliquen.
	\newblock {\em Granular media: between fluid and solid}.
	\newblock Cambridge University Press, 2013.
	
	\bibitem{11BZCB}
	Dapeng Bi, Jie Zhang, Bulbul Chakraborty, and Robert~P Behringer.
	\newblock Jamming by shear.
	\newblock {\em Nature}, 480(7377):355--358, 2011.
	
	\bibitem{16PMJ}
	Ivo~R Peters, Sayantan Majumdar, and Heinrich~M Jaeger.
	\newblock Direct observation of dynamic shear jamming in dense suspensions.
	\newblock {\em Nature}, 532(7598):214--217, 2016.
	
	\bibitem{18DJDZB}
	Dong Wang, Jie Ren, Joshua~A Dijksman, Hu~Zheng, and Robert~P Behringer.
	\newblock Microscopic origins of shear jamming for 2d frictional grains.
	\newblock {\em Physical review letters}, 120(20):208004, 2018.
	
	\bibitem{18BC}
	Robert~P Behringer and Bulbul Chakraborty.
	\newblock The physics of jamming for granular materials: a review.
	\newblock {\em Reports on Progress in Physics}, 82(1):012601, 2018.
	
	\bibitem{74Chen}
	HS~Chen and DE~Polk.
	\newblock Mechanical properties of ni fe based alloy glasses.
	\newblock {\em Journal of non-crystalline solids}, 15(2):174--178, 1974.
	
	\bibitem{79Argon}
	AS~Argon and HY~Kuo.
	\newblock Plastic flow in a disordered bubble raft (an analog of a metallic
	glass).
	\newblock {\em Materials science and Engineering}, 39(1):101--109, 1979.
	
	\bibitem{81Spaepen}
	Frans Spaepen.
	\newblock Defects in amorphous metals.
	\newblock {\em Les Houches lectures XXXV on physics of defects. North Holland:
		Amsterdam}, pages 136--74, 1981.
	
	\bibitem{01BLSLG}
	Lyderic Bocquet, W~Losert, D~Schalk, TC~Lubensky, and JP~Gollub.
	\newblock Granular shear flow dynamics and forces: Experiment and continuum
	theory.
	\newblock {\em Physical review E}, 65(1):011307, 2001.
	
	\bibitem{10Schall}
	Peter Schall and Martin Van~Hecke.
	\newblock Shear bands in matter with granularity.
	\newblock {\em Annual Review of Fluid Mechanics}, 42:67--88, 2010.
	
	\bibitem{13Greer}
	AL~Greer, YQ~Cheng, and E~Ma.
	\newblock Shear bands in metallic glasses.
	\newblock {\em Materials Science and Engineering: R: Reports}, 74(4):71--132,
	2013.
	
	\bibitem{07Tordesillas}
	A~Tordesillas.
	\newblock Force chain buckling, unjamming transitions and shear banding in
	dense granular assemblies.
	\newblock {\em Philosophical Magazine}, 87(32):4987--5016, 2007.
	
	\bibitem{08BB}
	Brian Utter and RP~Behringer.
	\newblock Experimental measures of affine and nonaffine deformation in granular
	shear.
	\newblock {\em Physical review letters}, 100(20):208302, 2008.
	
	\bibitem{zaccone2011approximate}
	Alessio Zaccone and Enzo Scossa-Romano.
	\newblock Approximate analytical description of the nonaffine response of
	amorphous solids.
	\newblock {\em Physical Review B}, 83(18):184205, 2011.
	
	\bibitem{14ZDB}
	Hu~Zheng, Joshua~A Dijksman, and Robert~P Behringer.
	\newblock Shear jamming in granular experiments without basal friction.
	\newblock {\em Europhysics Letters}, 107(3):34005, 2014.
	
	\bibitem{04ML}
	Craig Maloney and Ana{\"{e}}l Lema{\^{i}}tre.
	\newblock {Universal Breakdown of Elasticity at the Onset of Material Failure}.
	\newblock {\em Physi. Rev. Lett.}, 93(19):195501, nov 2004.
	
	\bibitem{06ML}
	Craig~E. Maloney and Ana\"el Lema\^{\i}tre.
	\newblock Amorphous systems in athermal, quasistatic shear.
	\newblock {\em Phys. Rev. E}, 74:016118, Jul 2006.
	
	\bibitem{98Falk}
	Michael~Lawrence Falk.
	\newblock {\em Deformation and fracture in amorphous solids}.
	\newblock University of California, Santa Barbara, 1998.
	
	\bibitem{13DHP}
	Ratul Dasgupta, H.~George~E. Hentschel, and Itamar Procaccia.
	\newblock Yield strain in shear banding amorphous solids.
	\newblock {\em Phys. Rev. E}, 87:022810, Feb 2013.
	
	\bibitem{20DeGiuli}
	Eric De~Giuli.
	\newblock Renormalization of elastic quadrupoles in amorphous solids.
	\newblock {\em Physical Review E}, 101(4):043002, 2020.
	
	\bibitem{FalkLanger98}
	M.~L. Falk and J.~S. Langer.
	\newblock Dynamics of viscoplastic deformation in amorphous solids.
	\newblock {\em Phys. Rev. E}, 57:7192--7205, Jun 1998.
	
	\bibitem{SollichLequeux97}
	Peter Sollich, Fran{\c{c}}ois Lequeux, Pascal H{\'e}braud, and Michael~E Cates.
	\newblock Rheology of soft glassy materials.
	\newblock {\em Physical review letters}, 78(10):2020, 1997.
	
	\bibitem{livne2023geometric}
	Noemie~S Livne, Amit Schiller, and Michael Moshe.
	\newblock Geometric theory of mechanical screening in two-dimensional solids.
	\newblock {\em Physical Review E}, 107(5):055004, 2023.
	
	\bibitem{14CKPUZ}
	Patrick Charbonneau, Jorge Kurchan, Giorgio Parisi, Pierfrancesco Urbani, and
	Francesco Zamponi.
	\newblock {Fractal free energy landscapes in structural glasses}.
	\newblock {\em Nature Communications}, 5:4725, apr 2014.
	
	\bibitem{sun2015granular}
	Qicheng Sun, Feng Jin, Guangqian Wang, Shixiong Song, and Guohua Zhang.
	\newblock On granular elasticity.
	\newblock {\em Scientific reports}, 5(1):9652, 2015.
	
	\bibitem{argon1979plastic}
	AS~Argon.
	\newblock Plastic deformation in metallic glasses.
	\newblock {\em Acta metallurgica}, 27(1):47--58, 1979.
	
	\bibitem{taub1980kinetics}
	AI~Taub and F~Spaepen.
	\newblock The kinetics of structural relaxation of a metallic glass.
	\newblock {\em Acta Metallurgica}, 28(12):1781--1788, 1980.
	
	\bibitem{nampoothiri2022tensor}
	Jishnu~N Nampoothiri, Michael D'Eon, Kabir Ramola, Bulbul Chakraborty, and
	Subhro Bhattacharjee.
	\newblock Tensor electromagnetism and emergent elasticity in jammed solids.
	\newblock {\em Physical Review E}, 106(6):065004, 2022.
	
	\bibitem{friedland1980geometric}
	L~Friedland and IB~Bernstein.
	\newblock Geometric optics in plasmas characterized by non-hermitian dielectric
	tensors.
	\newblock {\em Physical Review A}, 22(4):1680, 1980.
	
	\bibitem{avron1998odd}
	JE~Avron.
	\newblock Odd viscosity.
	\newblock {\em Journal of statistical physics}, 92:543--557, 1998.
	
	\bibitem{scheibner2020odd}
	Colin Scheibner, Anton Souslov, Debarghya Banerjee, Piotr Sur{\'o}wka,
	William~TM Irvine, and Vincenzo Vitelli.
	\newblock Odd elasticity.
	\newblock {\em Nature Physics}, 16(4):475--480, 2020.
	
	\bibitem{wang2020sheared}
	Dong Wang, Joshua~A Dijksman, Jonathan Bar{\'e}s, Jie Ren, and Hu~Zheng.
	\newblock Sheared amorphous packings display two separate particle transport
	mechanisms.
	\newblock {\em Physical Review Letters}, 125(13):138001, 2020.
	
	\bibitem{22MMPRSZ}
	Chandana Mondal, Michael Moshe, Itamar Procaccia, Saikat Roy, Jin Shang, and
	Jie Zhang.
	\newblock Experimental and numerical verification of anomalous screening theory
	in granular matter.
	\newblock {\em Chaos, Solitons \& Fractals}, 164:112609, 2022.
	
	\bibitem{22KMPS}
	Avanish Kumar, Michael Moshe, Itamar Procaccia, and Murari Singh.
	\newblock Anomalous elasticity in classical glass formers.
	\newblock {\em Physical Review E}, 106(1):015001, 2022.
	
	\bibitem{moshe2015geometry}
	Michael Moshe, Ido Levin, Hillel Aharoni, Raz Kupferman, and Eran Sharon.
	\newblock Geometry and mechanics of two-dimensional defects in amorphous
	materials.
	\newblock {\em Proceedings of the National Academy of Sciences},
	112(35):10873--10878, 2015.
	
	\bibitem{moshe2015elastic}
	Michael Moshe, Eran Sharon, and Raz Kupferman.
	\newblock Elastic interactions between two-dimensional geometric defects.
	\newblock {\em Physical Review E}, 92(6):062403, 2015.
	
	\bibitem{bar2020geometric}
	Yohai Bar-Sinai, Gabriele Librandi, Katia Bertoldi, and Michael Moshe.
	\newblock Geometric charges and nonlinear elasticity of two-dimensional elastic
	metamaterials.
	\newblock {\em Proceedings of the National Academy of Sciences},
	117(19):10195--10202, 2020.
	
	\bibitem{moshe2014isf}
	Michael Moshe, Eran Sharon, and Raz Kupferman.
	\newblock The plane stress state of residually stressed bodies: A stress
	function approach.
	\newblock {\em arXiv preprint arXiv:1409.6594}, 2014.
	
	\bibitem{22CMP}
	Harish Charan, Michael Moshe, and Itamar Procaccia.
	\newblock Anomalous elasticity and emergent dipole screening in
	three-dimensional amorphous solids.
	\newblock {\em Physical Review E}, 107(5):055005, 2023.
	
	\bibitem{de1999granular}
	Pierre-Gilles de~Gennes.
	\newblock Granular matter: a tentative view.
	\newblock {\em Reviews of modern physics}, 71(2):S374, 1999.
	
	\bibitem{alexander1998amorphous}
	Shlomo Alexander.
	\newblock Amorphous solids: their structure, lattice dynamics and elasticity.
	\newblock {\em Physics reports}, 296(2-4):65--236, 1998.
	
	\bibitem{howell1999stress}
	Daniel Howell, Robert~P Behringer, and Christian Veje.
	\newblock Stress fluctuations in a 2d granular couette experiment: A continuous
	transition.
	\newblock {\em Physical Review Letters}, 82(26):5241, 1999.
	
	\bibitem{veje1999kinematics}
	CT~Veje, Daniel~W Howell, and RP~Behringer.
	\newblock Kinematics of a two-dimensional granular couette experiment at the
	transition to shearing.
	\newblock {\em Physical Review E}, 59(1):739, 1999.
	
	\bibitem{ren13prl}
	J.~Ren, J.~A. Dijksman, and R.~P. Behringer.
	\newblock Reynolds pressure and relaxation in a sheared granular system.
	\newblock {\em Phys. Rev. Lett.}, 110(1):018302, 2013.
	
	\bibitem{wang20prl}
	Dong Wang, Joshua~A. Dijksman, Jonathan Bar\'es, Jie Ren, and Hu~Zheng.
	\newblock Sheared amorphous packings display two separate particle transport
	mechanisms.
	\newblock {\em Phys. Rev. Lett.}, 125:138001, Sep 2020.
	
	\bibitem{zadeh19_gm}
	A.~A. Zadeh, J.~Bar{\'e}s, T.~A. Brzinski, K.~E. Daniels, J.~A. Dijksman,
	N.~Docquier, H.~Everitt, J.~E. Kollmer, O.~Lantsoght, D.~Wang, M.~Workamp,
	Y.~Zhao, and H.~Zheng.
	\newblock Enlightening force chains: a review of photoelasticimetry in granular
	matter.
	\newblock {\em Granular Matter}, 21(83), 2019.
	
	\bibitem{howell97pg}
	D.~Howell and R.~P. Behringer.
	\newblock Fluctuations and dynamics for a two-dimensional sheared granular
	material.
	\newblock In R.~P. Behringer and J.~T. Jenkins, editors, {\em Powders and
		Grains 97}, pages 337--340. Taylor \& Francis, 1997.
	
	\bibitem{howell99prl}
	D.~Howell, R.~P. Behringer, and C.~Veje.
	\newblock Stress fluctuations in a 2d granular couette experiment: A continuous
	transition.
	\newblock {\em Phys. Rev. Lett.}, 82:5241--5244, 1999.
	
\end{thebibliography}


%

\end{document}